\documentclass[manuscript]{emulateapj}
\usepackage{amsmath}
\usepackage{amssymb}
\usepackage{natbib}
\usepackage{hyperref}
\usepackage{breakurl}

\newcommand{\mail}{lilirayhk@gmail.com}

\newcommand{\cm}{\,cm$^{-2}$}
\newcommand{\nh}{$N_\mathrm{H}$}

\shorttitle{The X-ray modulation of PSR~J2032+4127/MT91~213 during the Periastron Passage in 2017}
\shortauthors{Li et al., }

\begin{document}
\title{The X-ray modulation of PSR~J2032+4127/MT91~213 during the Periastron Passage in 2017}
\author{
K. L. Li\altaffilmark{1}, J. Takata\altaffilmark{2}, C. W. Ng\altaffilmark{3}, A. K. H. Kong\altaffilmark{4}, P. H. T. Tam\altaffilmark{5}, C. Y. Hui\altaffilmark{6}, and K. S. Cheng\altaffilmark{3}
}

\altaffiltext{1}{Department of Physics and Astronomy, Michigan State University, East Lansing, MI 48824, USA; \href{mailto:\mail}{\mail}}
\altaffiltext{2}{School of physics, Huazhong University of Science and Technology, Wuhan 430074, China}
\altaffiltext{3}{Department of Physics, The University of Hong Kong, Pokfulam Road, Hong Kong}
\altaffiltext{4}{Institute of Astronomy, National Tsing Hua University, Hsinchu 30013, Taiwan}
\altaffiltext{5}{Institute of Astronomy and Space Science, Sun Yat-Sen University, Guangzhou 510275, China}
\altaffiltext{6}{Department of Astronomy and Space Science, Chungnam National University, Daejeon 305-764, Republic of Korea}

\begin{abstract}
We present the \textit{Neil Gehrels Swift Observatory} (\textit{Swift}), \textit{Fermi Large Area Telescope} (\textit{Fermi}-LAT), and \textit{Karl G. Jansky Very Large Array} (VLA) observations of the $\gamma$-ray binary PSR J2032+4127/MT91 213, of which the periastron passage has just occurred in November 2017. In the \textit{Swift} X-ray light curve, the flux was steadily increasing before mid-October 2017, however, a sharp X-ray dip on a weekly time-scale is seen during the periastron passage, followed by a post-periastron X-ray flare lasting for $\sim20$ days. 
We suggest that the X-ray dip is caused by (i) an increase of the magnetization parameter at the shock, and (ii) the suppression due to the Doppler boosting effect. 
The 20-day post-periastron flare could be a consequence of the Be stellar disk passage by the pulsar. 
An orbital GeV modulation is also expected in our model, however, no significant variability is seen in the \textit{Fermi}-LAT light curve. 
We suspect that the GeV emission from the interaction between the binary's members is hidden behind the bright magnetospheric emission of the pulsar. Pulsar gating technique would be useful to remove the magnetospheric emission and recover the predicted GeV modulation, if an accurate radio timing solution over the periastron passage is provided in the future. 
\\
\end{abstract}
\keywords{X-rays: binaries --- pulsars: individual (PSR~J2032+4127) --- stars: individual (MT91~213) --- stars: winds, outflows}

\section{Introduction}
PSR~J2032+4127 is a young pulsar that has shown pulsations at a spin period of $P_s=143.2$~ms in both $\gamma$-rays \citep{2009Sci...325..840A} and radio \citep{2009ApJ...705....1C,2011ApJS..194...17R}. A subsequent timing study by \cite{2015MNRAS.451..581L} indicated that PSR~J2032+4127 is orbiting in a highly-eccentric orbit with the Be star, MT91 213 in the Cyg OB2 stellar association \citep{1991AJ....101.1408M}. Based on the latest timing solution published by the team, the binary has a very long orbital period of 45--50 years with an eccentricity of $e=0.94$--0.99 and the pulsar would have reached the periastron in November 2017 \citep{2017MNRAS.464.1211H}. \\

The binary PSR~J2032+4127/MT91~213 (J2032 hereafter) has been suggested to be a $\gamma$-ray binary: a subclass of high-mass X-ray binaries (HMXBs) whose the members show high-energy (HE; 0.1--100~GeV) and/or very-high-energy (VHE; $>100$~GeV) orbital modulations in their highly eccentric orbits (see, e.g., \citealt{2013A&ARv..21...64D}). While the pulsed emission of J2032 could be too bright to dominate over the possible HE modulation in GeV \citep{2017ApJ...836..241T}, VERITAS and MAGIC found that the TeV emission of J2032 increased by a factor of $\sim10$ from June/August to November 2017 \citep{2017ATel10810....1V,2017ATel10971....1M}. In addition to VHE, the X-ray emission of J2032 has been rapidly increasing in 2016--17 \citep{2017MNRAS.464.1211H,2017ApJ...843...85L}, and this pre-periastron X-ray enhancement has been commonly seen in other $\gamma$-ray binaries, e.g., PSR~B1259$-$63/LS~2883 \citep{2015ApJ...798L..26T,2015MNRAS.454.1358C}. 
\cite{2017ApJ...836..241T} proposed an intra-binary shock model, which involves an evolving pulsar wind magnetization and the Doppler Boosting effect, to explain the pre-periastron X-ray rise. Alternatively, \citealt{2018MNRAS.474L..22P} adopted an axisymmetric (no azimuthal dependence) stellar wind structure to explain the observed X-ray light curve. Besides the global increasding trend, \cite{2017ApJ...843...85L} found strong spectral variability on a monthly time-scale in X-rays, but the mechanism behind still remains unclear. \\

In this paper, we report the new \textit{Swift} observations taken during the periastron passage. \textit{Fermi}-LAT observations are also presented, however no significant $\gamma$-ray variability can be detected. We also discussed how such an X-ray modulation can be formed with the pulsar wind/stellar wind interaction model \citep{2017ApJ...836..241T}. 
Throughout the analysis, we assumed the periastron date to be 2017 Nov 12 (MJD 58069) as suggested by the \texttt{Model 2} in \cite{2017MNRAS.464.1211H}, although 2017 Nov 13 as the the periastron date was derived in some recent \textit{Astronomer's Telegrams} (e.g., \citealt{2017ATel10920....1C}). \\

\section{\textit{Swift} observations}
Since early-2016, \textit{Swift} has been intensively monitoring J2032, from a weekly cadence before mid-2017 to a daily cadence around the periastron in Nov 2017. As of 31 Jan 2018, 177 usable observations can be found in the \textit{Swift} public data archive (6 of them were taken before 2016). Most of them have exposure times between 1~ksec and 4~ksec, and a few have 5 ksec or more. \\

\subsection{XRT Data Reduction}
The \textit{Swift}/XRT products on-line tools\footnote{\url{http://www.swift.ac.uk/user_objects/index.php}} (\texttt{HEAsoft} v6.22 based) are used to build the light curve and the spectra used in this study \citep{2007A&A...469..379E,2009MNRAS.397.1177E}. Except for switching (i) the binning method to \texttt{Observation}, (ii) the centroid method to \texttt{Iterative}, and (iii) the minimum significance for a detection to 2, all default parameters were adopted. 
In addition, we manually subtracted the expected contribution from the three XRT-unresolved sources (i.e., $1.7\times10^{-3}$ cts/s) in the light curve to avoid an overestimation of the X-ray emission (see the detailed calculation in \citealt{2017ApJ...843...85L}). To compute the count rate to flux conversion factor, we extracted spectra using the observations from 2017-04-06 to 2017-11-12 (arbitrarily chosen) and fitted them with an absorbed power-law simultaneously (all the parameters are tiled except the normalizations). The best-fit parameters are \nh\ $=1.1\times10^{22}$\cm\ and $\Gamma=1.6$, which yield a conversion factor of $1.142\times10^{-10}$ erg cm$^{-2}$ cts$^{-1}$ (for unabsorbed flux in 0.3--10~keV). We also tried some other combinations of spectra and the conversion factor does not change significantly. Finally, although the XRT data qualities do not allow a good time-resolved spectral analysis over the periastron passage, we calculated the hardness ratio for each epoch to study the evolution of the X-ray color (i.e., $H/S$, where $S$ is the soft X-ray count rate in 0.3--1.5~keV and $H$ is the hard X-ray count rate in 1.5--10~keV; Figure~\ref{xlight}). \\

\subsection{UVOT Data Reduction}
\texttt{HEAsoft} (v6.22) with the UVOT \texttt{CALDB} (v201709221) was used to reduce the UVOT observations. All six UVOT filters have been used (one filter per observation in most of the cases), however, the $v$- and $b$-band light curves are under-sampled and are therefore not discussed in this paper. \\

The \textit{Swift}-specific \texttt{FTOOLS}, \texttt{uvotmaghist}, was used to extract the UV light curves using aperture photometry. The source aperture was chosen to be a 3\arcsec\ radius circular region, which is the optimal size for the UVOT data\footnote{\url{https://swift.gsfc.nasa.gov/analysis/threads/uvot_thread_aperture.html}}. Two bright sources are fairly close to J2032. To accurately accounting for the contamination from them, we used a 3\arcsec\ circular region as the background region, at a position that the distances from the two nearby sources are the same as the distances between the nearby sources and J2032. 
Although the $u$-band images are slightly overexposed as MT91~213 is very bright with $m_v=11.95$~mag \citep{2003AJ....125.2531R}, the flags \texttt{saturated=0} from \texttt{uvotmaghist} suggest that the measurements are still usable. \\


\section{\textit{Fermi}-LAT observations}
\label{section:fermilc}
To obtain the $\gamma$-ray long-term light curve of J2032, we downloaded the \textit{Fermi} ``Pass 8 Source'' LAT data (instrumental response function: ``P8R2\_SOURCE\_V6'') from the \textit{Fermi} Science Support Center (FSSC)\footnote{\url{http://fermi.gsfc.nasa.gov/ssc/}} with the criteria of (i) energy from 100 MeV to 500 GeV, and (ii) time from 2008-08-04 to 2018-01-22. A region of interest (ROI) was chosen to be $20^\circ \times 20^\circ$ square centered at the epoch J2000 position of the source: $(\textrm{R.A.},\textrm{Dec.})=(20^{\textrm{h}} 32^{\textrm{m}} 14\fs35,+41^\circ 26^\prime 48\farcs8)$ (i.e., the LAT position of J2032 in the 3FGL catalog; \citealt{ac15}).  In addition, all the data observed at zenith angles greater than $90^\circ$ were excluded to avoid contamination from the Earth's albedo. 
All the data reduction and analysis processes were performed using the \textit{Fermi} Science Tools package version v10r0p5. 

\begin{figure*}
 \centering
 \includegraphics[width=1\textwidth]{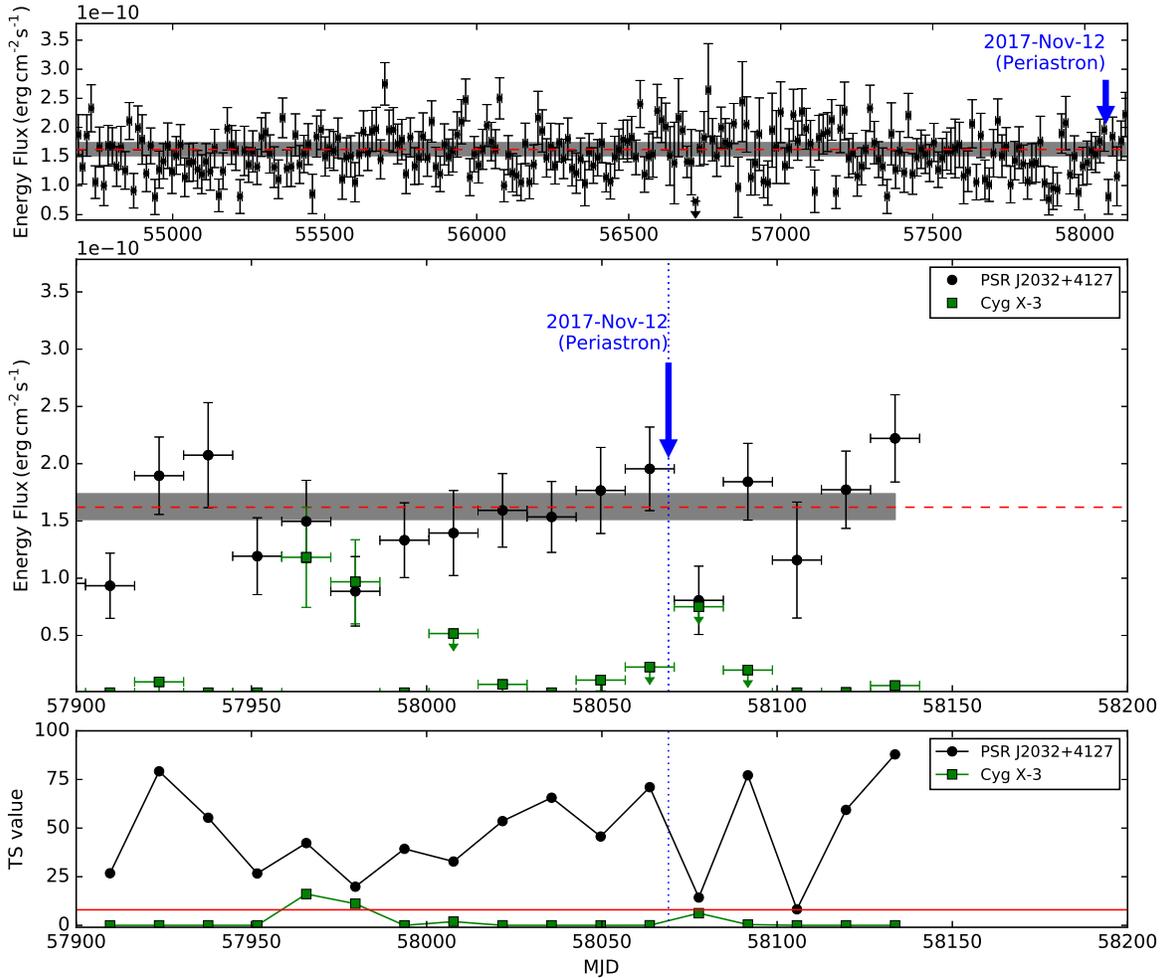}
 \caption{(Top) The \textit{Fermi} $\gamma$-ray light curve of J2032 during 2008-08-04 to 2018-01-22 in the energy range 100 MeV -- 500 GeV. (Middle) A zoom-in version of the light curve showing the energy flux of J2032 (circle) and the nearby Cyg~X-3 (square) from 2017-05-27 to 2018-01-22. In both light curves, the blue arrow shows the time of the periastron passage, while the red dashed line represents the average flux level of the source from the global binned likelihood analysis during 2008-08-04 to 2017-10-19 with the corresponding uncertainty indicated by the gray shaded band. (Bottom) The test-statistic (TS) values of J2032 and Cyg~X-3. The red solid line indicates the TS threshold of 8, under which an 95\% upper limit (triangle) is determined in the above light curves. }
 \label{fig:J2032_fermi_lc}
\end{figure*}

We first used the \texttt{gtlike} tool to model the average emission from the background sources between 2008-08-04 to 2017-10-19 with a maximum likelihood optimization technique (i.e., the binned likelihood analysis). The source model includes (i) all the 3FGL cataloged sources within $25^\circ$ from the center of the ROI (gll\_psc\_v16.fit; \citealt{ac15}), (ii) the galactic diffuse emission (gll\_iem\_v06), and the isotropic diffuse emission (iso\_P8R2\_SOURCE\_V6\_v06), and (iii) the nearby microquasar Cygnus X-3 \citep{bo13} located $<0.5^\circ$ away from J2032 (a simple power-law spectral model was assumed). For those 3FGL sources that are non-variable and located $>6^\circ$ away from the ROI center, all the spectral parameters were fixed to their listed values in the 3FGL. There are also four extended sources in the source model: Gamma Cygni, Cygnus Cocoon, HB 21 and Cygnus Loop, which were modeled by the extended source templates obtained from the FSSC. Our target J2032, known as 3FGL~J2032.2+4126 in the 3FGL catalog, was described by a power-law with simple exponential cutoff in the source model, 
\begin{equation}
 \frac{\textrm{d}N}{\textrm{d}E}=N_0 \left(\frac{E}{E_0}\right)^{-\Gamma} \textrm{exp} \left(-\frac{E}{E_C}\right),
 \label{equation:plsec}
\end{equation}
where $N_0$ is the normalization constant, $E_0$ is the scale factor of energy in MeV, $\Gamma$ is the spectral power-law index, and $E_C$ is the cut-off energy in MeV. From the binned likelihood analysis, the best-fit parameters of J2032 during 2008-08-04 to 2017-10-19 are $N_0=(1.66 \pm 0.05)\times 10^{-11}$, $\Gamma=-1.39 \pm 0.04$, and $E_C=(4500 \pm 249)$~MeV. \\

Next, we construct a new model from the above version by fixing all the spectral parameters to their global best-fit values, except the normalizations. The new model was then used to compute a long-term 2-week binned light curve of J2032 with the binned likelihood method, which is shown in Figure~\ref{fig:J2032_fermi_lc}. In addition, the light curve of Cyg~X-3 is shown for comparison. As shown in the bottom panel of the figure, Cyg~X-3 was mostly undetected at $\sim 3\sigma$ significance level around the periastron passage, indicating that the observed variability of J2032, however weak, is unlikely to be related to the transient nature of Cyg~X-3. It is also important to note that the light curve shows the total energy fluxes observed at the position of J2032, which is the sum of the pulsar's magnetospheric emission (which is the dominant component) and the possible contribution from the interaction between the binary's members. \\

\section{VLA observation}
\begin{figure*}
 \centering
 \includegraphics[width=\textwidth]{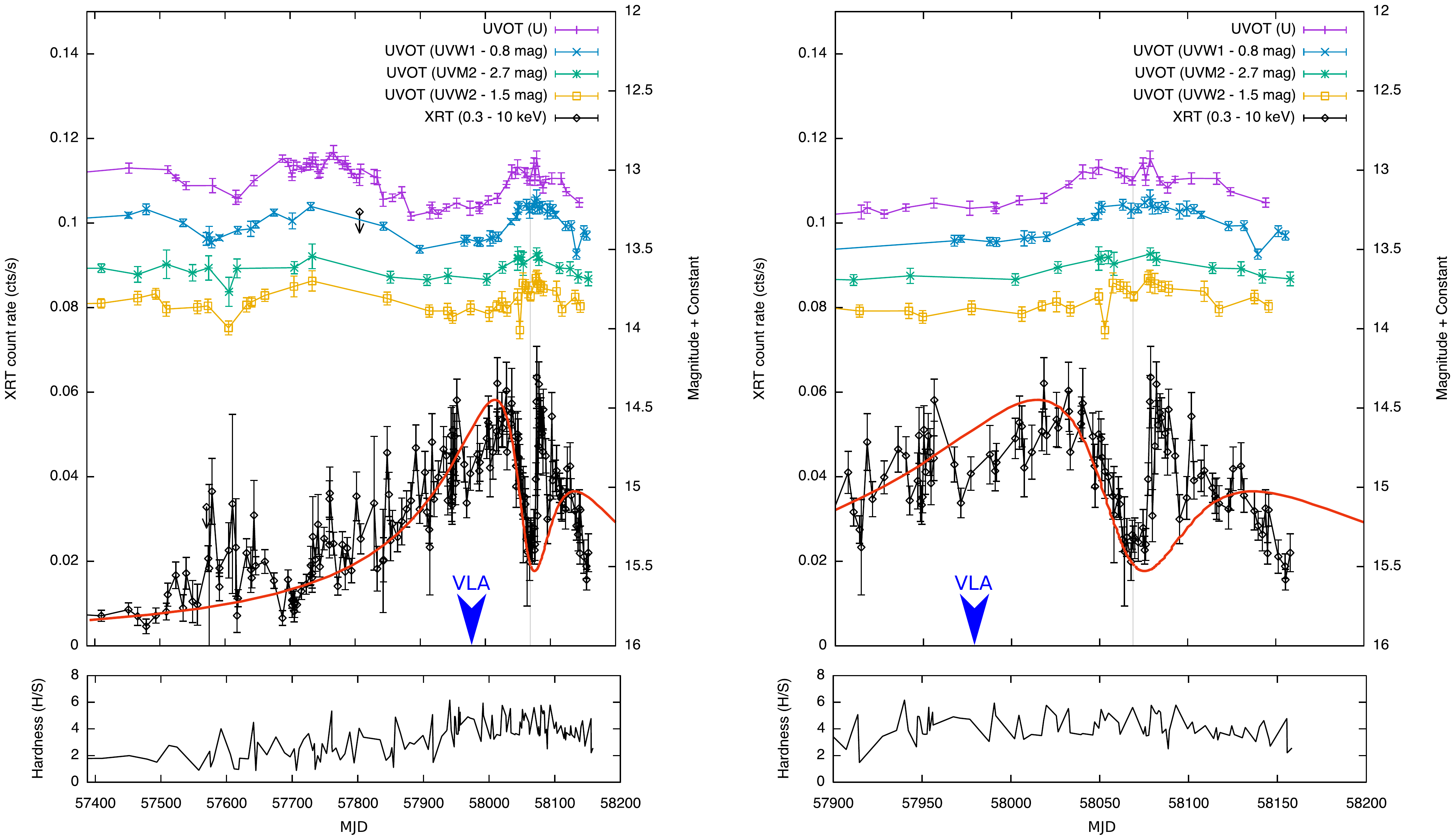}
 \caption{(Left) The 0.3--10.0 keV and the UVOT light curves of J2032 (from 2016 to 2018 January; 95\% upper limits for non-detection epochs) with the hardness ratio evolution (i.e., $H/S$, where $S=0.3$--1.5~keV and $H=1.5$--10~keV) in the lower panel. (Right) The zoom-in version for the periastron passage. Vega magnitude system is used for the UVOT light curve. For the hardness ratio plots, only those data points with uncertainties less than 2 are shown. 
The periastron date and the VLA observing date (2017-08-14 or MJD 57979) were indicated by a grey vertical line and a blue arrow, respectively. 
Finally, the red solid line shows the model light curve with $\alpha=2$ for the radial distribution of the magnetization, and 40\% of the speed of the post-shocked flow assumed in \cite{2017ApJ...836..241T}. A more detailed description for the model can be found in \cite{2017ApJ...836..241T}.} 
\label{xlight}
 \end{figure*}

We observed J2032 with the VLA at 3 GHz (2--4~GHz; S-band) in the C configuration on 2017 August 14, from 05:10:40 to 05:43:20 UTC (the observing date was marked in Figure \ref{xlight}), under a Director Discretionary Time (DDT). Standard data reduction procedures were performed using the CASA software package (v4.7.2). J2032 was clearly detected with an average flux of $S_3=0.10$ mJy~beam$^{-1}$ (background rms noise: 0.014 mJy~beam$^{-1}$). The obtained flux is well consistent with the previous measurement taken in 2009 (i.e., $S_2=0.12$ mJy at 2 GHz; \citealt{2009ApJ...705....1C}), implying that the system did not evolve much at least in August (3 months before the periastron passage). It is worth noting that the VLA observation was taken near the first peak of the X-ray light curve (Figure \ref{xlight}). While the X-rays increased by a factor of 4 from 2016 to 2017, the radio remains roughly the same. \\

\section{Discussions}

\subsection{X-ray Modulation}
The observed X-ray flux of J2032 showed a rapid increase in 2013--2016, and then it abruptly began to decrease
in 2017 October just before the periastron passage in early November. In addition, the X-ray hardness ratio likely increased (i.e., harder) when the pulsar was passing the periastron, possibly because of the higher \nh, hence the heavier soft X-ray absorption, near the Be star. Despite the small scale fluctuation and/or flare-like
behaviors observed soon after the periastron passage that could be related to the interactions between the pulsar and the clumpy stellar
wind and/or the Be stellar disk, the global trend of the observed X-ray light curve can be explained by the evolution of the magnetization of
the pulsar wind and the Doppler boosting effect of the shocked pulsar wind. The magentization of the pulsar wind is
defined by the ratio of the magnetic energy to the kinetic energy 
\begin{equation}
 \sigma=\frac{B^2}{4\pi \Gamma_{\rm PW}N_{\rm PW}\,m_ec^2},
 \end{equation}
where $B$, $\Gamma_{\rm PW}$, and $N_{\rm PW}$ are the magnetic field, Lonrentz factor, and number density of the cold-relativistic pulsar wind,
respectively, at the region between the pulsar and the shock. It has been a long standing problem of how the magnetization evolves with distance from $\sigma\gg1$ at the light cylinder to $\sigma<1$ at the interstellar shock of the pulsar wind nebula (see, e.g., \citealt{1984ApJ...283..694K,1990ApJ...349..538C,2001ApJ...547..437L}). \cite{2011ApJ...729..104K} and \cite{2017PhRvL.119u1101K} suggested that the evolution can be described by $\sigma\propto r^{-1}$ (here $r$ means the radial distance from the pulsar) for the regions faraway from the light cylinder in the absence of the magnetic dissipation. The evolution can be steeper if there is a magnetic dissipation in the pulsar wind. \\

We have modeled the X-ray modulation of J2032 over the periastron passage by assuming a radial dependency of the magnetization as $\sigma\propto r^{-\alpha}$, where $\alpha$ is in the range of 1 -- 3 \citep{2017ApJ...836..241T}.
In the model, the X-rays are dominated by the synchrotron emission from the post-shocked pulsar wind with the shock geometry calculated based on $\eta\sim 0.02$, which is the momentum ratio of the spin-down power of the pulsar ($L_{\rm sd}\sim 1.7\times 10^{35}\,{\rm erg~s^{-1}}$) to the stellar wind. The Doppler boosting effect due to the finite velocity of the shocked pulsar wind were also considered, assuming a constant bulk velocity of the post-shock pulsar along the shock-cone that was calculated based on the jump condition of a perpendicular magnetohydrodynamics (MHD) shock \citep{1984ApJ...283..694K}. \\

By comparing the model to the XRT light curve before 2017, the rapid X-ray flux increase in 2013--2016 implies a radial evolution with an index of $\alpha=2-3$ \citep{2017ApJ...836..241T}. In addition, a rapid decrease in X-rays around the periastron was predicted, based on (i) an increase of the magnetization parameter at the shock (i.e., $\sigma>1$), and (ii) the suppression due to the Doppler boosting effect. The Doppler boosting effect would also make the observed X-ray modulation asymmetric about the periastron date as the viewing angle changes. These features have all been seen in the observed \textit{Swift}/XRT light curve shown in Figure \ref{xlight}, although the post-periastron X-ray emission predicted in \cite{2017ApJ...836..241T} was slightly underestimated (see Figure~16 with $\alpha=2$ in the reference). This likely implies an overestimation of the speed of the post-shocked pulsar wind flow, and hence the X-ray flux suppression due to the Doppler boosting effect. We therefore reduced the assumed speed of the post-shock flow down to 40\% and the resultant model light curve matches the general trend of the XRT light curve reasonably well (Figure \ref{xlight}). \\

It is not straightforward to understand the flare-like X-ray structure observed around MJD 58080--58100 in the pulsar wind/stellar wind interaction model. Alternatively, this X-ray enhancement could be caused by the pulsar and Be stellar disk interaction, which abruptly changes the shock structure. The radius of the shock (from the pulsar) induced by the interaction can be determined by
\begin{equation}
 r_s=\left(\frac{L_{\rm sd}}{2\pi \rho_{d} v_{r}^2c}\right)^{1/2},
 \end{equation}
where $\rho_{d}$ is the disk mass density at the pulsar's position and $v_{r}$ is the relative velocity between the pulsar and the disk rotation.
If the scale height of the Be stellar disk at the pulsar's position is larger than the shock radius, the disk can confine
most of the pulsar wind, which could lead to an X-ray flux enhancement as suggested in \cite{2012ApJ...750...70T}. In \cite{2017ApJ...836..241T}, we discussed that if the base density of the Be stellar disk is larger than $\rho_0>10^{-10}\,{\rm g~cm^{-3}}$ and the pulsar/Be stellar disk interaction occurs at the periastron passage, the Be stellar disk can make a cavity of the pulsar wind around the pulsar, and enhance the X-ray emission. The flare-like X-ray enhancement lasting for about 20~days, which is a much longer period compared to the time-scale needed for the pulsar to cross the Be stellar disk (i.e., $t_{c}\sim H/v_{p}\sim 2$ days, where $H\sim 0.1$~AU is
the scale height of the Be stellar disk at the pulsar orbit and $v_{p}\sim 10^7\,{\rm cm~s^{-1}}$ is the pulsar's orbital velocity). However, some disk matter could pile up in front of the pulsar when it is passing through the disk. This phenomenon has been shown possible in the 3D smoothed particle hydrodynamic (SPH) simulation by \cite{2012ApJ...750...70T}. This piled-up disk matter will influence the shock structure, and hence the X-ray emission, until it is dispersed by the pressure of the nearby gas. The time-scale of the dispersion can be estimated as
$t_d\sim H/c_s\sim 20$ days, where $c_s\sim$ 10 km~s$^{-1}$ is the sound speed of the Be stellar disk \citep{2011PASJ...63..893O}, and it is well consistent with the time-scale of the X-ray flare. \\

\subsection{Possible Orbital Modulation in UV}
\label{sec:uv}
In the UVOT light curve, there is a clear UV brightening on a time-scale of about $100$ days right at the periastron passage. Unfortunately, the UV brightening is strongly contaminated by MT~91/213 (see the strong intrinsic variability of the Be star before MJD 57900 in Figure \ref{xlight}), and therefore not much information can be extracted. In fact, it is entirely possible that the UV brightening is totally unrelated to the periastron passage, but merely a time coincidence to the intrinsic brightness change of the Be star. \\

In \cite{2017ApJ...836..241T}, we discussed the possibility that if the density of the Be stellar disk is sufficiently high, some matter of the disk could be captured by the pulsar during the periastron passage and a short-lived (e.g., weeks) UV-emitting accretion disk can be formed around the pulsar. However, the accretion disk would be very faint compared to the Be star/disk (a factor of $\gg10$ lower), and therefore the UV band (below 0.01 keV) will still be completely dominated by the emission from the Be star/disk (see Figure 22 in \citealt{2012ApJ...750...70T}). Obviously, the faint UV emission from the aforementioned accretion disk is not comparable to the UV brightening, which has a relatively high amplitude of $\sim$0.2~mag (about 20\% of the Be star/disk). We therefore conclude that the enhanced UV emission is unlikely from the accretion disk. \\

\subsection{Gamma-Ray Light Curve}
We did not see any significant GeV modulation in the \textit{Fermi}-LAT light curve, which is not totally unexpected given the strong contamination from the bright pulsed $\gamma$-ray emission of PSR~J2032+4127 as we have mentioned. Performing pulsar gating could remove the unwanted pulsar's contribution, and hence recover the GeV modulation due to the pulsar/stellar winds interactions, if an accurate ephemeris of the pulsar during the periastron passage is provided by radio timing observations. \\

Yet, there was a very marginal $\gamma$-ray flux drop (about 50\% of the mean flux) observed right after the periastron passage (the middle panel of Figure \ref{fig:J2032_fermi_lc}). While the drop is totally consistent with the statistical fluctuations seen in other epochs of the light curve indicating that the drop is insignificant, we note that a possible accretion flow discussed in \S\ref{sec:uv} can indeed shut down the $\gamma$-ray emission from the pulsar's magnetosphere (see, e.g., \citealt{2017ApJ...836..241T}), resulting in a similar light curve feature. In this scenario, the radio pulsation should have been shut down either. Radio observations taken in the post-periastron epoch would be very useful to test the idea. \\

\begin{acknowledgements}
Support for this work was partially provided by the National Aeronautics and Space Administration through Chandra Award Number GO7-18036X issued by the Chandra X-ray Observatory Center, which is operated by the Smithsonian Astrophysical Observatory for and on behalf of the National Aeronautics Space Administration under contract NAS8-03060. 
JT is supported by NSFC grants of Chinese Government under 11573010 and U1631103. PHT is supported by the NSFC grant 11633007. Both are supported by NSFC grant 11661161010. 
AKHK is supported by the Ministry of Science and Technology of the Republic of China (Taiwan) through grants 105-2119-M-007-028-MY3 and 105-2112-M-007-033-MY2. 
CYH is supported by the National Research Foundation of Korea through grant 2016R1A5A1013277. 
KSC are supported by GRF grant under 17302315. 
We acknowledge the use of public data from the \textit{Swift} data archive. 
We also acknowledge the use of data and software facilities from the FSSC, managed by the HEASARC at the Goddard Space Flight Center. 
The National Radio Astronomy Observatory is a facility of the National Science Foundation operated under cooperative agreement by Associated Universities, Inc.\\
\end{acknowledgements}
\textit{Facilities}: \facility{Swift, Fermi, VLA}

\bibliography{j2032b}

\end{document}